\documentclass[amsmath,amssymb,twocolumn,prl,floatfix,showpacs,nofootinbib]{revtex4-2}
\usepackage{amsmath}
\usepackage{graphicx}
\usepackage{subfigure}
\usepackage{epstopdf}
\usepackage{color}
\usepackage{multirow}
\usepackage{setspace}
\usepackage{overpic}
\usepackage{amssymb}
\usepackage[colorlinks,urlcolor=blue, citecolor=blue,linkcolor=blue] {hyperref}
\usepackage{lineno}
\usepackage{bm}
\usepackage{rotating}
\usepackage[utf8]{inputenc}
\let\oldequation\equation
\let\oldendequation\endequation
\renewenvironment{equation}
 {\linenomathNonumbers\oldequation}
 {\oldendequation\endlinenomath}

\begin{document}

\title{\bf \boldmath
First Measurement of the Absolute Branching Fraction of $\Lambda \to p \mu^- \bar{\nu}_{\mu}$
}

\author{
\begin{small}
\begin{center}
M.~Ablikim$^{1}$, M.~N.~Achasov$^{10,b}$, P.~Adlarson$^{67}$, S. ~Ahmed$^{15}$, M.~Albrecht$^{4}$, R.~Aliberti$^{28}$, A.~Amoroso$^{66A,66C}$, M.~R.~An$^{32}$, Q.~An$^{63,49}$, X.~H.~Bai$^{57}$
% ...
\\
\vspace{0.2cm}
(BESIII Collaboration)\\
\vspace{0.2cm} {\it
$^{1}$ Institute of High Energy Physics, Beijing 100049, People's Republic of China\\
% ...
$^{4}$ Bochum Ruhr-University, D-44780 Bochum, Germany\\
% ...
$^{10}$ G.I. Budker Institute of Nuclear Physics SB RAS (BINP), Novosibirsk 630090, Russia\\
% ...
$^{15}$ Helmholtz Institute Mainz, Staudinger Weg 18, D-55099 Mainz, Germany\\
% ...
$^{28}$ Johannes Gutenberg University of Mainz, Johann-Joachim-Becher-Weg 45, D-55099 Mainz, Germany\\
% ...
$^{32}$ Liaoning Normal University, Dalian 116029, People's Republic of China\\
% ...
$^{49}$ State Key Laboratory of Particle Detection and Electronics, Beijing 100049, Hefei 230026, People's Republic of China\\
% ...
$^{57}$ University of Jinan, Jinan 250022, People's Republic of China\\
% ...
$^{63}$ University of Science and Technology of China, Hefei 230026, People's Republic of China\\
% ...
$^{66}$ University of Turin and INFN, (A)University of Turin, I-10125, Turin, Italy; (B)University of Eastern Piedmont, I-15121, Alessandria, Italy; (C)INFN, I-10125, Turin, Italy\\
$^{67}$ Uppsala University, Box 516, SE-75120 Uppsala, Sweden\\
% ...
$^{b}$ Also at the Novosibirsk State University, Novosibirsk, 630090, Russia\\
% ...
}
\end{center}
\end{small}
}

\begin{abstract}
The absolute branching fraction of $\Lambda \to p \mu^-
\bar{\nu}_{\mu}$ is reported for the first time based on an $e^+e^-$
annihilation sample of ten billion $J/\psi$ events collected with the
BESIII detector at $\sqrt{s}=3.097$ GeV.  The branching fraction is
determined to be ${\mathcal B}(\Lambda \to p \mu^- \bar{\nu}_{\mu}) =
[1.48\pm0.21(\rm stat) \pm 0.08(\rm syst)]\times 10^{-4}$, which is a
significant improvement in precision over the previous indirect
measurements.  Combining this result with the world average of
${\mathcal B}(\Lambda \to p e^- \bar{\nu}_{e})$, we obtain the ratio,
$\frac{\Gamma(\Lambda \to p \mu^- \bar{\nu}_{\mu})}{\Gamma(\Lambda \to
  p e^- \bar{\nu}_{e})}$, to be $0.178 \pm 0.028$, which agrees with
the standard model prediction assuming lepton flavor universality.
The asymmetry of the branching fractions of $\Lambda \to p \mu^-
\bar{\nu}_{\mu}$ and $\bar{\Lambda} \to \bar{p} \mu^+ \nu_{\mu}$ is
also determined, and no evidence for $CP$ violation is found.
\end{abstract}

\maketitle

\oddsidemargin  -0.2cm
\evensidemargin -0.2cm


The Standard Model (SM) of particle physics provides precise predictions for the
properties and interactions of fundamental particles, 
which have been confirmed by numerous experimental results
(\emph{e.g.} the discovery of the Higgs boson~\cite{higgs_atlas,higgs_cms}).
However, recently there have been indications of tensions between theory and experiment, in
particular in the lepton sector~\cite{pdg2020}.

Semileptonic (SL) hyperon decays provide a benchmark to test 
the SM and complement direct searches for physics beyond the SM, 
especially for muonic modes which are very sensitive to
non-standard scalar and tensor contributions~\cite{sm_r}.  In the SM,
the SL hyperon decays are described by $SU(3)$ flavor 
symmetry, which enables systematic expansions and accurate predictions
with a simplified dependence on hadronic form factors~\cite{sm_r}.
Therefore, a comparison of the branching fraction~(BF) ${\mathcal
  B}(\Lambda \to p \mu^- \bar{\nu}_{\mu})$ between its experimental
measurement and its SM expectation provides an important
probe of physics beyond the SM.

Lepton flavor universality (LFU), which is an accidental feature of
the SM~\cite{lfu_review}, has been tested in recent years using a
variety of different probes, and there are hits for a possible violation of LFU
in semileptonic $b$-quark decays.  The measurements are obtained from
experiments at the B-factories (BaBar~\cite{old_result_1,old_result_2}
and
Belle~\cite{old_result_3,old_result_4,old_result_5,old_result_5_1}),
as well as at the LHC
(LHCb)~\cite{old_result_6,old_result_7,old_result_8,old_result_9}.
According to the results from the Heavy Flavor Averaging Group, a
combined discrepancy at the level of three standard deviations is
observed in $b \to c \ell \bar{\nu}_{\ell}$ decays~\cite{HFAG}.
A similar comprehensive analysis of exotic effects in $s
\to u$ transitions has not yet been done, especially for SL hyperon decays,
which can be denoted as $B_1 \to B_2 \ell^- \bar{\nu}_{\ell}$.  For
the SL hyperon decays, the LFU test observable is the ratio between decay rates of
the semimuonic decay and the semielectronic decay, $R^{\mu e} \equiv
\frac{\Gamma(B_1 \to B_2 \mu^- \bar{\nu}_{\mu})}{\Gamma(B_1 \to B_2
  e^- \bar{\nu}_{e})}$, which is not only sensitive to LFU violation
but is also linearly sensitive to the contributions of
(pseudo-)scalar and tensor operators~\cite{sm_r}.

In theory, working at next-to-leading order, the LFU test observable
$R^{\mu e}$ of $\Lambda \to p$ decay is predicted to be
$0.153\pm0.008$~\cite{sm_r}, while the current experimental
measurement is $0.189\pm0.041$~\cite{pdg2020}.  The large experimental
uncertainty is dominated by the BF ${\mathcal
  B}(\Lambda \to p \mu^- \bar{\nu}_{\mu})$. So far, experimental
information for ${\mathcal B}(\Lambda \to p \mu^- \bar{\nu}_{\mu})$
has only come from fixed-target experiments~\cite{old_result_FBC,old_result_RVUE,old_result_HBC_1971,old_result_HBC_1972}, 
which were performed about fifty years ago. The most
precise measurement was performed in 1972~\cite{old_result_HBC_1972} and
was reported as a relative BF $\frac{\Gamma(\Lambda \to p \mu^-
  \bar{\nu}_{\mu})}{\Gamma(\Lambda \to {\rm N}\pi)}=(1.4\pm0.5)\times
10^{-4}$ based on fourteen signal events which were selected from about
0.6 million bubble chamber pictures.  With the current level of
precision, the experimental $R^{\mu e}$ result agrees with the SM
prediction.  A more accurate measurement of ${\mathcal B}(\Lambda \to
p \mu^- \bar{\nu}_{\mu})$ will provide a more stringent test of LFU.

In addition, it is possible to test for $CP$ violation, which has been
observed in $K$~\cite{cp_1} and $B$ meson decays~\cite{cp_2,cp_3} and
in 2019 in neutral charm meson decays~\cite{cp_4}. However, all
effects observed so far of $CP$ violation in particle decays cannot
explain the observed matter-antimatter asymmetry in the Universe. This
motivates further searches for new sources of $CP$ violation, which
has not yet been observed in the decays of any baryon.  Within the SM,
$CP$ violation for down-type quarks ($s$ or $b$) is expected to be
larger than for up-type quark ($c$)~\cite{cp_sm}, which motivates us
to search for $CP$ violation in hyperon decays.  In 2019, the BESIII
collaboration reported the most precise direct test of $CP$ violation
in $\Lambda$ hyperon nonleptonic decays $\Lambda \to p \pi^{-}$ and
$\bar{\Lambda} \to \bar{p} \pi^{+},\bar{n}\pi^{0}$~\cite{cp_5}.  In
comparison, no search for $CP$ violation in SL hyperon decays has yet been
reported. Hence, a search for $CP$ violation in SL hyperon decays offers
complementary information in the hyperon sector.

In this Letter, we report the first measurement of the
absolute BF $\mathcal{B}(\Lambda \to p \mu^- \bar{\nu}_{\mu})$, by
analyzing $\Lambda\bar{\Lambda}$ hyperon pairs in ten billion $J/\psi$
meson decay events collected with the BESIII detector at
$\sqrt{s}=3.097$ GeV.  We use the double-tag (DT) technique~\cite{DT},
which provides a clean and straightforward BF measurement without
requiring knowledge of the total number of $\Lambda\bar{\Lambda}$
events produced.  Based on the measured absolute branching fraction,
$\mathcal{B}(\Lambda \to p \mu^- \bar{\nu}_{\mu})$, $R^{\mu e}$
for $\Lambda$ semileptonic decays is determined.  In addition, the
$CP$ asymmetry of $\Lambda \to p \mu^- \bar{\nu}_{\mu}$ and
$\bar{\Lambda} \to \bar{p} \mu^+ \nu_{\mu}$ is also presented for the
first time.


Details about the design and performance of the BESIII detector are
given in Refs.~\cite{BESIII,whitepaper}.  Simulated data samples
produced with a {\sc Geant4}-based~\cite{geant4} Monte Carlo (MC)
software, which includes the geometric description of the BESIII
detector and the detector response, are used to determine the
detection efficiencies and to estimate backgrounds. The simulation
includes the beam energy spread and initial state radiation in the
$e^+e^-$ annihilations modeled with the generator {\sc
  kkmc}~\cite{kkmc}.  For the simulations of both of the decays $\Lambda \to p
\mu^- \bar{\nu}_{\mu}$ and $\Lambda \to p e^- \bar{\nu}_{e}$, we use
the form factors of $\Lambda \to p e^- \bar{\nu}_{e}$ obtained from experimental measurements, 
which are summarized in Ref.~\cite{FF}.  The generator constructed in
Ref.~\cite{cp_5} is used to simulate the dominant background $\Lambda
\to p \pi^-$ decay.  An `inclusive' MC sample of generic events includes both the
production of the $J/\psi$ resonance and the continuum processes
incorporated in {\sc kkmc}~\cite{kkmc}.  The known decay modes are
modeled with {\sc evtgen}~\cite{evtgen} using BFs taken from the
Particle Data Group~\cite{pdg2020}, and the remaining unknown
charmonium decays are modeled with {\sc
  lundcharm}~\cite{lundcharm}. Final state radiation from charged
final state particles is incorporated with {\sc photos}~\cite{photos}.



Using the DT technique, we obtain the BF by reconstructing signal
$\Lambda \to p \mu^- \bar{\nu}_{\mu}$ decays in events with
$\bar{\Lambda}$ decays reconstructed in its dominant hadronic decay
mode, $\bar{\Lambda} \to \bar{p} \pi^+$.  If a $\bar{\Lambda}$ hyperon
is found, it is referred to as a single-tag (ST) candidate.  An event in
which a signal $\Lambda$ decay and an ST $\bar{\Lambda}$ are
simultaneously found is referred as a DT event. The BF of the signal
decay is given by

\begin{equation}
\label{eq_br}
{\mathcal B_{\rm sig}}= \frac{N_{\rm DT}/\epsilon_{\rm DT}}{N_{\rm ST}/\epsilon_{\rm ST}},
\end{equation}
where $N_{\rm DT}$ is the DT yield, $\epsilon_{\rm DT}$ is the DT
selection efficiency, and $N_{\rm ST}$ and $\epsilon_{\rm ST}$ are the
ST yield and the ST selection efficiency. Throughout this paper, 
charge-conjugated channels are always implied.

Good charged tracks detected in the main drift chamber (MDC) must
satisfy $|\cos \theta| < 0.93$, where $\theta$ is the polar angle
with respect to the $z$ axis, which is the axis of the MDC.  Events
with at least two good charged tracks are selected.  Combinations of
any pair of oppositely-charged tracks are assigned as ST
$\bar{\Lambda}$ candidates without imposing further particle
identification (PID) criteria.  The pairs are constrained to originate from
a common vertex by requiring the $\chi^2$ of the vertex fit to be less
than 100.  The decay length of the $\bar{\Lambda}$ candidate is
required to be greater than twice the vertex resolution away from the
interaction point. At least one $\bar{\Lambda}$ hyperon is required to
be reconstructed successfully via the vertex fits.  The tagged
$\bar{\Lambda}$ hyperons are selected using two variables, the energy
difference
\begin{equation}
\label{def_delE}
\Delta E_{\rm tag} \equiv E_{\bar{\Lambda}}-E_{\rm beam},
\end{equation}
and the beam-constrained mass
\begin{equation}
\label{def_mbc}
M_{\rm BC}^{\rm tag}c^2 \equiv \sqrt{E^{2}_{\rm beam}-|\vec{p}_{\bar{\Lambda}}c|^{2}},
\end{equation}
where $E_{\rm beam}$ is the beam energy, and $\vec{p}_{\bar{\Lambda}}$ and $E_{\bar{\Lambda}}$
are the momentum and the energy of the $\bar{\Lambda}$ candidate in the $e^+e^-$ rest frame.
If there are multiple combinations, the one giving the minimum $|\Delta E_{\rm tag}|$ is retained for further analysis.
The tagged $\bar{\Lambda}$ are required to satisfy
$\Delta E_{\rm tag}\in[-17,\, 13]$\,MeV.

\begin{figure}[htp]
  \centering
\includegraphics[width=1\linewidth]{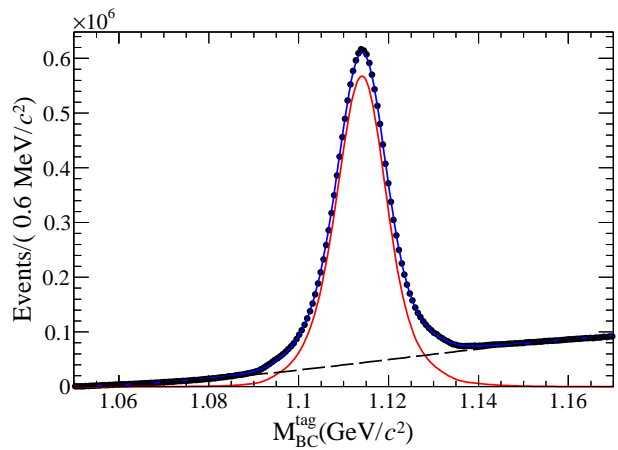}
  \caption{\small
Fit to the $M_{\rm BC}^{\rm tag}$ distribution of the ST
$\bar{\Lambda} + c.c.$ candidates.
Data are shown as dots with error bars.
The solid blue, solid red and dashed black curves are the fit result, signal shape and the background shape, respectively.}
\label{fig_STfit}
\end{figure}

The yield of ST $\bar{\Lambda}$ hyperons is obtained from a maximum
likelihood fit to the $M_{\rm BC}^{\rm tag}$ distribution of the
accepted ST candidates, where we use the MC-simulated signal shape
convolved with a double-Gaussian resolution function to represent the
signal shape and a third order Chebyshev function to describe the
backgrounds.  The signal yield is estimated in the mass region
$[1.089, 1.143]$~GeV/$c^2$.  The fit result is shown in
Fig.~\ref{fig_STfit}, and the total ST $\bar{\Lambda} +c.c.$ yield is
$N_{\rm ST}=14,609,800\pm7,117(\rm stat)$.

\begin{table*}[htp]
\centering
\caption{\label{tab_all}
\small
The $N_{\rm ST}$, $N_{\rm DT}$, $\epsilon_{\rm ST}$, $\epsilon_{\rm DT}$ and the obtained BFs.
The uncertainties are statistical only.}
\begin{tabular*}{\hsize}{@{}@{\extracolsep{\fill}}cccccc@{}}
\hline\hline
Decay mode&							$N_{\rm ST}~(\times\,10^{3})$&		$N_{\rm DT}$& 	$\epsilon_{\rm ST}$\,(\%)&	$\epsilon_{\rm DT}$\,(\%)&	${\mathcal B_{\rm sig}}~(\times\,10^{-4})$\\
\hline
$\Lambda \to p \mu^- \bar{\nu}_{\mu} +c.c.$	&$14,609.8\pm7.1$&			$64\pm9$&		$55.36\pm0.05$&			$1.65\pm0.01$&			$1.48\pm0.21$ \\
$\Lambda \to p \mu^- \bar{\nu}_{\mu}$		&$7,385.9\pm5.1$&				$31\pm7$&		$55.21\pm0.06$&			$1.64\pm0.01$&			$1.43\pm0.30$ \\
$\bar{\Lambda} \to \bar{p} \mu^+ \nu_{\mu}$	&$7,391.0\pm5.0$&				$33\pm6$&		$55.50\pm0.08$&			$1.66\pm0.01$&			$1.49\pm0.29$ \\
\hline\hline
\end{tabular*}
\end{table*}


Candidate events for $\Lambda \to p \mu^- \bar{\nu}_{\mu}$ decays are
selected from the remaining tracks recoiling against the ST
$\bar{\Lambda}$ candidates. We require the total number of all good
charged tracks to be four ($N_{\rm Track}= 4$) with the criteria for
additional good charged tracks the same as those used in the ST
selection.  We further identify a charged track as a $\mu^-$ by
requiring the PID likelihoods calculated by combining the MDC ionization
energy loss, time-of-flight and electromagnetic calorimeter
information satisfy $\mathcal{L}_{\mu} > 0.001$ and $\mathcal{L}_{\mu}
> \mathcal{L}_{e}$, where the $\mathcal{L}_{\mu}$ and
$\mathcal{L}_{e}$ are likelihoods for the muon and electron
hypotheses, respectively. The other track is assumed to be a proton.
As the neutrino is not detected, we employ the kinematic variable
\begin{equation}
U_{\rm miss} \equiv E_{\rm miss}-c|\vec{p}_{\rm miss}|
\end{equation}
to obtain information on the neutrino, where $E_{\rm
miss}$ and $\vec{p}_{\rm miss}$ are the missing energy and momentum
carried by the neutrino, respectively.
$E_{\rm miss}$ is calculated by
 \begin{equation}
 E_{\rm miss}=E_{\rm beam}-E_{p}-E_{\mu^-},
 \end{equation}
 where $E_{p}$ and $E_{\mu^-}$ are the measured energies of $p$ and
 $\mu^-$, respectively.  We use the magnitude of the constrained
 $\Lambda$ momentum to calculate $p_{\rm miss}$
 \begin{equation}
   p_{\rm miss}=|\vec{p}_{\Lambda}-{\vec p}_{p}-{\vec p}_{\mu^-}|,
 \end{equation}
 where $\vec{p}_{\Lambda}$, ${\vec p}_{p}$ and ${\vec p}_{\mu^-}$ are
 the momenta of $\Lambda$, $p$ and $\mu^-$, respectively, in which
 $\vec{p}_{\Lambda}$ is given by
 \begin{equation}
   \vec{p}_{\Lambda}=-\frac{\vec{p}_{\bar{\Lambda}}}{c|\vec{p}_{\bar{\Lambda}}|}\sqrt{E_{\rm beam}^{\rm 2}-m_{\Lambda}^{\rm 2}c^4},
 \end{equation}
where $m_{\Lambda}$ is the nominal $\Lambda$ mass.
For signal events, $U_{\rm miss}$ is expected to peak around zero.

For the accepted signal candidates of $\Lambda \to p \mu^-
\bar{\nu}_{\mu}$ decay, there is still background from the dominant
hadronic decay $\Lambda \to p \pi^-$, because of misidentification
between $\mu^-$ and $\pi^-$ and $\pi^-$ decay which leads to $\Lambda
\to p \pi^{-}\to p \mu^- \bar{\nu}_{\mu}$ background.  To suppress
this background, we first impose a four-constraint energy momentum
conservation (4C-fit) kinematic fit with the $J/\psi \to\Lambda\bar{\Lambda}$
hypothesis.  Before the 4C-fit, a $\Lambda$ is reconstructed based on
the $p \pi^-$ hypothesis to obtain the momentum vector of the $\Lambda$. The
$\chi^2$ of the 4C-fit is required to be larger than twenty.  Second,
for this background, the mass recoiling against $\bar{\Lambda}p$,
\emph{i.e.} $M_{\bar{\Lambda}p}^{\rm recoil}$, is expected to be the
$\pi^-$ mass.  Therefore, we require the signal
candidates satisfy $M_{\bar{\Lambda}p}^{\rm recoil}>0.170~ {\rm
  GeV}/c^{2}$.  This requirement can effectively suppress the $\Lambda \to p
\pi^{-}$ background, resulting in the relative signal
efficiency being 34 times larger than that of the background.  Third,
after the 4C-fit, if we assign the $\pi^-$ mass to $\mu^-$ candidates
when calculating the invariant mass of $p \mu^-$,
\emph{i.e.} $M_{p\mu(4C)}^{\rm sig}$, we can eliminate background by only
retaining the events with $M_{p\mu(4C)}^{\rm sig} \in [1.075,\, 1.100
]~ {\rm GeV}/c^{2}$.  To verify the reliability of these requirements,
ten cross checks varying the criteria above and below the nominal
requirements have been performed using the method reported in
Ref.~\cite{xi_patrik}.

The inclusive MC sample is analyzed using TopoAna~\cite{topology} to study
potential backgrounds.  After imposing the above selection criteria,
there is no peaking background in the signal region, and the dominant
backgrounds are $\Lambda \to p \pi^-$ and $\Lambda \to p e^-
\bar{\nu}_{e}$ decays that are included in the determination of the
signal yield.  For the potential backgrounds that include an extra photon, $J/\psi \to \gamma
\Lambda \bar{\Lambda}$ and $\Lambda \to p \pi^- \gamma$ decays, which
are studied with corresponding exclusive MC simulation, the $J/\psi \to \gamma
\Lambda \bar{\Lambda}$ decay background is negligible.  The $\Lambda
\to p \pi^- \gamma$ decay background is small but will be taken into
consideration as a systematic uncertainty.

\begin{figure}[htp]
  \centering
\includegraphics[width=1\linewidth]{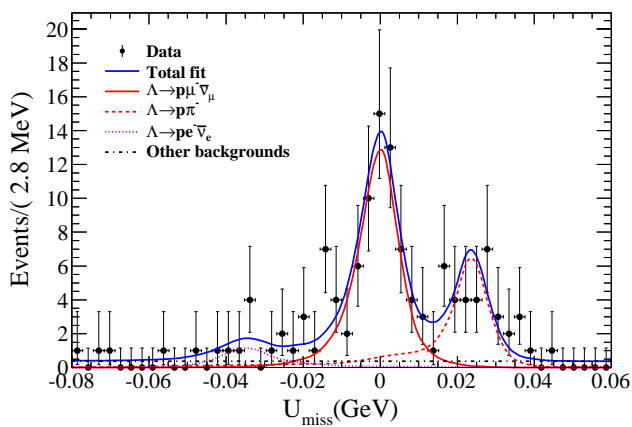}
  \caption{\small Fit to the $U_{\rm miss}$ distribution of the DT
    candidates. Data are shown as dots with error bars.  The solid blue and
    red curves are the fit result and signal shape,
    respectively. The dashed red and dotted violet curves are
    background shapes for $\Lambda \to p \pi^-$ and $\Lambda \to p e^-
    \bar{\nu}_{e}$ decays, respectively. The dash-dotted black curve
    represents the other backgrounds.  }
\label{fig_DTfit}
\end{figure}

To determine the signal yield, an unbinned extended maximum likelihood
fit is performed to the $U_{\rm miss}$ distribution.  The signal is
modeled by the MC-simulated signal shape convolved with a Gaussian
resolution function to account for imperfect simulation of the detector
resolution.  The main backgrounds are modeled by the MC-simulated
shapes obtained from the exclusive MC samples.  Other backgrounds are
described by a first-order polynomial.  The parameters of the
Gaussian, the first-order polynomial, and all yields are left free in
the fit.  The fit to the data is shown in Fig.~\ref{fig_DTfit}.
The numbers of $N_{\rm ST}$, $\epsilon_{\rm ST}$, $N_{\rm DT}$,
$\epsilon_{\rm DT}$ and the BF of $\Lambda \to p \mu^-
\bar{\nu}_{\mu} +c.c.$ are summarized in the first row of
Table~\ref{tab_all}.

The systematic uncertainties due to the requirements for $N_{\rm
  Track}= 4$ (2.71\%), $\Lambda$ reconstruction through the vertex fit
(0.05\%), the 4C-fit (0.57\%) and the $M_{\bar{\Lambda}p}^{\rm
  recoil}>0.170~ {\rm GeV}/c^{2}$ and $M_{p\mu(4C)}^{\rm sig} \in
[1.075,\, 1.100 ]~ {\rm GeV}/c^{2}$ (1.04\%) are studied with the
control sample $J/\psi \to \Lambda(\to p \pi^{-}) \bar{\Lambda}(\to
\bar{p} \pi^{+})$ using the method reported in Ref.~\cite{cp_5}. For
the simulation of the signal MC model (2.80\%), it is estimated by
varying the input values of form factors~\cite{FF} by one standard
deviation.  Other sources of systematic uncertainty include the
following items: the MC statistics (0.01\%); the proton tracking
(1.00\%), muon tracking (1.00\%) and the muon PID (2.00\%), which are
cited from Refs.~\cite{p_tracking, sys_muon}; the fits to the $U_{\rm
  miss}$ (1.87\%) and $M_{\rm BC}^{\rm tag}$ (2.17\%) distributions
estimated by using alternative fit procedures, \emph{i.e.} changing the
signal and background shapes for both of these fits and changing the
bin size for the fit to the $M_{\rm BC}^{\rm tag}$ distribution.  For
the fit to $U_{\rm miss}$, the signal shape is changed by removing
the Gaussian resolution function, and the background shapes are
changed in three ways.  First, we convolve the background shapes with
the Gaussian resolution function which is the same as the one for the
signal shape.  Then, the $\Lambda \to p \pi^- \gamma$ MC-simulated
shape is added.  Finally, we change the input parameters~\cite{cp_5}
by one standard deviation to determine the $\Lambda \to p \pi^-$
MC-simulated shape.  The total systematic uncertainty is estimated to
be 5.55\% by adding all these uncertainties in quadrature.

Finally, we obtain the BF, ${\mathcal B}(\Lambda \to p \mu^-
\bar{\nu}_{\mu}) = (1.48\pm0.21\pm0.08)\times10^{-4}$, where the first
uncertainty is statistical and the second is systematic. Combining with the
well-measured BF of the decay $\Lambda \to p e^- \bar{\nu}_{e}$,
${\mathcal B}(\Lambda \to p e^- \bar{\nu}_{e})=(8.32\pm0.14)\times
10^{-4}$~\cite{pdg2020}, we determine the ratio $R^{\mu e} \equiv
\frac{\Gamma(\Lambda \to p \mu^- \bar{\nu}_{\mu})}{\Gamma(\Lambda \to
  p e^- \bar{\nu}_{e})}$ to be $R^{\mu e}=0.178 \pm 0.028$.  This
result is consistent within uncertainties with the value
$0.153\pm0.008$ that is expected from LFU in the SM~\cite{sm_r}.


The BFs of the charge-conjugated decays $\Lambda \to p \mu^-
\bar{\nu}_{\mu}$ and $\bar{\Lambda} \to \bar{p} \mu^+ \nu_{\mu}$,
${\mathcal B}_{\Lambda \to p \mu^- \bar{\nu}_{\mu}}$ and ${\mathcal
  B}_{\bar{\Lambda} \to \bar{p} \mu^+ \nu_{\mu}}$, are measured
separately.  The asymmetry of these two BFs is determined as
\begin{equation}
{{\mathcal A}_{CP}} \equiv \frac{{\mathcal B}_{\Lambda \to p \mu^- \bar{\nu}_{\mu}}-{\mathcal B}_{\bar{\Lambda} \to \bar{p} \mu^+ \nu_{\mu}}}
{{\mathcal B}_{\Lambda \to p \mu^- \bar{\nu}_{\mu}}+{\mathcal B}_{\bar{\Lambda} \to \bar{p} \mu^+ \nu_{\mu}}}.
\end{equation}
The corresponding $N_{\rm ST}$, $N_{\rm DT}$, $\epsilon_{\rm ST}$,
$\epsilon_{\rm DT}$ and the BFs are summarized in the last two rows of
Table~\ref{tab_all}.  The asymmetry is determined to be ${\mathcal
  A}_{CP} = 0.02\pm0.14(\rm stat)\pm0.02(\rm syst)$, where the
systematic uncertainties of $N_{\rm Track}= 4$,
$\Lambda$ reconstruction through the vertex fit, the 4C-fit, the
$M_{\bar{\Lambda}p}^{\rm recoil}>0.170~ {\rm GeV}/c^{2}$, the
$M_{p\mu(4C)}^{\rm sig} \in [1.075,\, 1.100 ]~ {\rm GeV}/c^{2}$ and
the signal MC model cancel.  Other systematic uncertainties are
estimated separately as above.  No evidence for $CP$ violation is
found.


In summary, using ten billion $J/\psi$ decay events collected with the
BESIII detector at $\sqrt{s}=3.097$ GeV, the semileptonic hyperon decay $\Lambda \to p \mu^-
\bar{\nu}_{\mu}$ is studied at a collider experiment for the first
time.  Based on the double-tag method, we report the first measurement of the
absolute BF of $\Lambda \to p \mu^- \bar{\nu}_{\mu}$ as
${\mathcal B}(\Lambda \to p \mu^- \bar{\nu}_{\mu}) = [1.48\pm0.21(\rm
  stat) \pm 0.08(\rm syst)]\times 10^{-4}$ which improves the
precision of the world average value significantly.  The
BF is consistent with theoretical predictions that incorporate quark
SU(3) flavor symmetry without symmetry breaking~\cite{wangruming}, and
predictions based on the factorization of the contribution of valence
quarks and chiral effects~\cite{amand}.

Using the well-measured branching fraction of the decay $\Lambda \to p
e^- \bar{\nu}_{e}$, we determine the ratio $R^{\mu e} \equiv
\frac{\Gamma(\Lambda \to p \mu^- \bar{\nu}_{\mu})}{\Gamma(\Lambda \to
  p e^- \bar{\nu}_{e})}$ to be $R^{\mu e}=0.178 \pm 0.028$ which is in
  agreement with the previous results but is the most precise to date.
  The $R^{\mu e}$ result agrees with LFU, and the higher precision can
  aid in the study of the (pseudo-)scalar and tensor operator
  contributions in theory~\cite{sm_r}.  The asymmetry of the BFs of
  charge-conjugated decays $\Lambda \to p \mu^- \bar{\nu}_{\mu}$ and
  $\bar{\Lambda} \to \bar{p} \mu^+ \nu_{\mu}$ is also determined. No
  evidence for $CP$ violation is found.

\acknowledgments
The BESIII collaboration thanks the staff of BEPCII and the IHEP computing center for their strong support. This work is supported in part by National Key R\&D Program of China under Contracts Nos. 2020YFA0406300, 2020YFA0406400; National Natural Science Foundation of China (NSFC) under Contracts Nos. 11805037, 11625523, 11635010, 11735014, 11822506, 11835012, 11935015, 11935016, 11935018, 11961141012, 12022510, 12025502, 12035009, 12035013, 12061131003; the Chinese Academy of Sciences (CAS) Large-Scale Scientific Facility Program; Joint Large-Scale Scientific Facility Funds of the NSFC and CAS under Contracts Nos. U1832121, U1732263, U1832207; CAS Key Research Program of Frontier Sciences under Contract No. QYZDJ-SSW-SLH040; 100 Talents Program of CAS; INPAC and Shanghai Key Laboratory for Particle Physics and Cosmology; ERC under Contract No. 758462; European Union Horizon 2020 research and innovation programme under Contract No. Marie Sklodowska-Curie grant agreement No 894790; German Research Foundation DFG under Contracts Nos. 443159800, Collaborative Research Center CRC 1044, FOR 2359, GRK 214; Istituto Nazionale di Fisica Nucleare, Italy; Ministry of Development of Turkey under Contract No. DPT2006K-120470; National Science and Technology fund; Olle Engkvist Foundation under Contract No. 200-0605; STFC (United Kingdom); The Knut and Alice Wallenberg Foundation (Sweden) under Contract No. 2016.0157; The Royal Society, UK under Contracts Nos. DH140054, DH160214; The Swedish Research Council; U. S. Department of Energy under Contracts Nos. DE-FG02-05ER41374, DE-SC-0012069.


\end{document}